\documentclass{aastex}
\usepackage{emulateapj5}
\usepackage{apjfonts}
\usepackage{epsf}

\slugcomment{Astronomical Journal, vol. 125 (June Issue, 2003), in press}

\shorttitle{A limit relation between black hole mass and H$\beta$ width}
\shortauthors{Wang}

\def\dotm{\dot{m}}
\def\dotM{\dot{M}}
\def\dotMedd{\dot{M}_{\rm Edd}}

\def\rblr{R_{\rm BLR}}

\begin{document}

\title{A Limit Relation between Black Hole Mass and H$\beta$ Width: Testing
Super-Eddington Accretion in Active Galactic Nuclei}

\author{Jian-Min Wang\altaffilmark{1,2}}

\affil{$^1$Institut f\"ur Astronomie und Astrophysik, Abt. Astronomie,
Universit\"at T\"ubingen, Sand 1, D-72076 T\"ubingen, Germany} 

\affil{$^2$Laboratory for High Energy Astrophysics,
Institute of High Energy Physics, CAS,
Beijing 100039, P.R. China }

\begin{abstract}
In this paper, we show that there is a limit relation between the black 
hole mass ($M_{\rm BH}$) and the width at the half maximum ($\upsilon_{\rm FWHM}$)
of H$\beta$ for active galactic nuclei (AGNs) with 
super-Eddington accretion rates. When a black hole has a super-Eddington 
accretion rate, the empirical relation of reverberation mapping
has two possible ways. First, it reduces to a relation between the black 
hole mass and the size of the broad line region due to the photon trapping
effects inside the accretion disk.  For the Kaspi et al.'s empirical 
reverberation relation, we get the limit relation as 
$M_{\rm BH}=(2.9 - 12.6)\times 10^6M_{\odot}
\left(\upsilon_{\rm FWHM}/10^3{\rm km~s^{-1}}\right)^{6.67}$,
called as the Eddington limit. Second, the Eddington limit luminosity
will be relaxed if the trapped photons can escape from the magnetized 
super-Eddington accretion disk via the photon bubble instability, and 
the size of the broad line region will be enlarged 
according to the empirical reverberation relation, leading to a relatively 
narrow width of H$\beta$. We call this the Begelman limit. 

Super-Eddington accretions in a sample composed of 164 AGNs have been
searched by this limit relation.  We find that most of them are well confined 
by the Eddington limit relation, namely most of the objects in the sample have
sub-Eddington accretion rates, but there are a handful of objects
locate between the Eddington and Begelman limit lines, they may be candidates 
of super-Eddington accretors in a hybrid structure of photon trapping and 
photon bubble instability.  The maximum width of H$\beta$ is in the reange 
of $(3.0 - 3.8)\times 10^3$~km~s$^{-1}$ for the maximum mass black holes with 
super-Eddington accretion rates among AGNs.  We suggest that the 
FWHM(H$\beta) - M_{\rm BH}$ relation is more reliable and 
convenient to test whether a source is super-Eddington and useful to probe 
the structure of the super-Eddington accretion process.  
\end{abstract}


\section{Introduction}
Accretion onto a supermassive black hole is regarded as the energy sources of AGNs
(Rees 1984).
Energy output from accretion disks strongly depends on their structures and radiation
efficiency, which is mainly controlled by the 
Eddington ratio, defined by the dimenssionless accretion rate
$\dotm=\dotM/\dot{M}_{\rm Edd}$, where 
$\dotMedd=1.39\times 10^{18}\eta_{-1}^{-1}\left(M_{\rm BH}/M_{\odot}\right)$g~s$^{-1}$,
$\dot{M}$ is the accretion rate, $M_{\rm BH}$ the black hole mass
and the accretion efficiency $\eta=0.1\eta_{-1}$ (Chen et al. 1997, Frank, King \& Raine 2002). 
It has been realized that many observable features of AGNs could 
be driven by $\dot{m}$ (Boroson \& Green 1992,
Boroson 2002), however, it is not easy to estimate $\dot{m}$ from observations.

We need a more reliable and convenient criteria based on theory to test whether AGNs have 
super-Eddington accretion rates.
Much attention has been given to estimate the black hole masses in 
AGNs and inactive galaxies (Kormendy \& Gebhardt 
2001) so that the black hole mass becomes an easily "measured" quantity. There 
are currently four fundamental ways to reliably estimate the mass of the black 
holes by: 1) reverberation mapping methods for AGNs (Wandel, Peterson 
\& Malkan 1999, Kaspi et al. 2000), 2) the empirical relation of the reverberation 
mapping (Netzer 2003, Vestergaard 2002), 3) relation between $M_{\rm R}-M_{\rm BH}$ 
(Kormendy \& Richstone 1995, Ho 1999, McLure \& Dunlop 2001), where
$M_{\rm R}$ is the absolute $R$-band magnitudes of the host galaxies, 4) the relation
$\sigma-M_{\rm BH}$ (Ferrarese \& Merrit 2000, Gebhardt et al. 2000, Tremaine et al.
2002), including the method from the fundamental plan, where $\sigma$ is the 
dispersion velocity. Employing the empirical relation
of reverberation mapping, one can conveniently obtain the masses of the black holes
in a large sample spanning a wide range of redshift (Netzer 2003). 
With these estimations, the Eddington ratio can be estimated
via $L/L_{\rm Edd}=\xi L_{5100}/L_{\rm Edd}$, where $\xi=5 - 9$ (Kaspi et al.
2000, Netzer 2003).  It has been found that some AGNs may have  
super-Eddington accretion rates due to $L/L_{\rm Edd}>1$. They consist of some 
narrow line Seyfert 1 galaxies and quasars (Collin \& Hur\'e 2001, Collin 
et al. 2002, King \& Puchnarewicz 2002,
Wang \& Netzer 2003). One of the most prominent features is the strong
soft X-ray humps (King 2002, King \& Puchnarewicz 2002, Wang \& Netzer 2003).

The structure of the super-Eddington accretion disk remains open. Begelman
(1978) and Begelman \& Meier (1982)
suggested that a photon trapping process may happen in a super-Eddington
accretion. The classical model of the slim accretion disk ($\dot{m}<50 - 100$)
based on the vertical
averaged equations advocated by Abramowicz et al (1988) removes the thermal
instability of the radiation pressure-dominated region 
since the advection as an efficient cooling
dominates over the diffusive cooling from the disk surface (Chen \& Taam 1993).
The detail treatment of the photon trapping inside the slim disk has been done
by Ohsuga et al. (2002), who showed the maximum luminosity is about Eddington 
luminosity. This is roughly in agreement with Wang \& Zhou (1999). On the other 
hand, Gammie (1998) showed that
a {\it magnetized} accretion disk may have photon bubble instability in the
radiation pressure-dominated region of the standard accretion disk model (Shakura
\& Sunyaev 1973). Such an instability in the super-Eddington accretion disks should
be enhanced since the photons are trapped inside the disks and the radiation pressure 
is much higher than that in Shakura \& Sunyaev's disk.
Begelman (2002) suggested that the inhomogeneous distribution of 
the accreted matter will be formed due to the photon bubble instability and the
maximum luminosity will exceed Eddington luminosity by a factor of one
hundred for galactic mass of the black hole. Both
the spectra and the behavior of variabilities from 
slim disk (Wang et al. 1999; Wang \& Netzer 2003; Mineshige et al. 
2000) are not sufficiently understood as well as the roles of the photon bubble 
instability (Begelman 2002).
It may be difficult to directly test the structure of the super-Eddington accretion 
disk in an individual objects, but statistic test may provide invaluable information.

We show in this paper that $L/L_{\rm Edd}=\xi L_{5100}/L_{\rm Edd}$  
can not properly represent the accretion disk rates  in AGNs, 
particularly for those with super-Eddington accretion rates. 
We suggest a limit relation between the black hole mass and H$\beta$ 
width based on the theoretical models to test whether the accretion disks in AGNs 
have super-Eddington accretion rates. These limit relations are useful 
to probe the structures of the disks with super-Eddington rates.

\section{The Eddington and Begelman constraints}
\subsection{Size of the broad line region}
Measuring the response of the broad emission line to the variations of the
continuum can provide the distance of the clouds from the central black hole
in AGNs. Kaspi et al. (2000) provided the results of the reverberation
mapping in a sample consisting of 34 AGNs and found a strong
correlation between the size of the broad line region ($R_{\rm BLR}$) and the 
luminosity at 5100{\AA}, namely, the empirical reverberation relation 
\begin{equation}
\rblr={\cal R}_0\left(\frac{\lambda L_{\lambda}}{10^{44}{\rm erg~s^{-1}}}\right)^{\beta}~
            {\rm lt-days},
\end{equation}
where $\beta=0.7$, ${\cal R}_0=32.9$ and 
$\lambda L_{\lambda}$ is the continuum luminosity at wavelength $\lambda=5100$\AA. 
This relation has been calibrated by Vestergaard (2002) using the bivariate correlated 
errors and intrinsic scatter algorithm. The specific values of ${\cal R}_0$ and 
$\beta$ are in debate (see Vestergaard 2002, Netzer 2003). A constant ionization
parameter $U$, defined by $U=L_{\rm ion}/4\pi R^2 N_e$, leads to a relation
$R_{\rm BLR}\propto L_{\rm ion}^{0.5}$ (Wandel, Peterson \& Malkan 1999),
where $N_e$ is the number density
of electrons in ionized medium and $L_{\rm ion}$ is the ionizing luminosity. This is
supported by the argument that the outer boundary of the BLR is determined by 
dust sublimation radius (Netzer \& Laor 1993). 
We keep the uncertainty of the index $\beta$ in mind, however it does not
influence the existence of the limit relation between $M_{\rm BH}$ and H$\beta$
width derived in the present paper. 

For a given black hole, the continuum luminosity from an accretion disk
increases with the accretion rate (but below the Eddington rate) and
hence an increase of the BLR size according to the empirical reverberation relation
(equation 1).  This implies that a high accretion rate
leads to a relatively narrow width of H$\beta$ for a sub-Eddington accretion disk in
AGN  if the assumption that the broad line emitting-clouds
are virialized in the black hole potential works (see equation 2). 
When the accretion rate exceeds the Eddington rate, the
radiated luminosity from the disk very weakly depends on the accretion rate
($L_{\rm disk}\propto \log \dot{M}$), but linearly proportional to the
mass of the black hole in the context of the slim disk model
(Abramowicz et al. 1988, Wang \& Zhou 1999; Mineshige et al. 2000;
Ohsuga et al. 2002). Thus the radiated luminosity is {\it saturated},
leading to a {\it constant} BLR size independent of the accretion rate for a fixed 
black hole. We call this the Eddington size and the Eddington width of the emission 
line accordingly. For AGNs with super-Eddington accretion rates,
the Eddington width of H$\beta$ directly reflects
the potential of the black holes. In the context of the photon trapping disk
a relation of $M_{\rm BH}$ and FWHM(H$\beta$) is then expected.

However, it has been argued that magnetized disks with dominance of
radiation pressure are suffering from photon bubble instability (Gammie 1998).
The disks become so inhomogeneous on scales much smaller than the radiation pressure
scale height that the radiation will be squeezed by gravity out of overdense region
enhanced by the magnetic tension while the underdense regions become tenuous due to the
acceleration of radiation force. Some of the trapped photons may be liberated from the 
disk via this instability.
Thus the Eddington limit may be relaxed by this way.
Begelman (2001, 2002) investigated the photon bubble instability in a super-Eddington
accretion disk and found the maximum radiated luminosity 
$L_{\rm max}\approx 300 M_{\rm BH,6}^{1/5}L_{\rm Edd}$, where 
$M_{\rm BH,6}=M_{\rm BH}/10^6M_{\odot}$, keeping a
geometrically thin shape,  when the accretion rate reaches its maximum. 
We call this the Begelman limit.  It is not clear
what the radiated luminosity is before $\dot{M}$ reaches its maximum as well as
the emergent spectrum from such  disks. If the photon bubble instability works  
as a mechanism liberating the trapped photons in AGN super-Eddington accretion 
disks, the BLR size will be larger than the Eddington size, resulting in a narrower
width of H$\beta$ than the Eddington width. Thus the different structure of the 
super-Eddington accretion 
disk should lead to different limit relations between FWHM(H$\beta$) and $M_{\rm BH}$.
This lends us a possibility to statistically test the structure of the super-Eddington
accretion disk  based on the available data of FWHM(H$\beta$) and $M_{\rm BH}$.

\subsection{A limit relation between FWHM(H$\beta$) and $M_{\rm BH}$}
The broad line H$\beta$ emitting clouds are virialized in the gravitational potential 
of the supermassive black hole, the mass of the black hole is expressed by the 
FWHM(H$\beta$)
\begin{equation}
M_{\rm BH}=1.46\times 10^5f^2\left(\frac{R_{\rm BLR}}{{\rm lt-days}}\right)
           \left(\frac{\upsilon_{\rm FWHM}}{10^3{\rm km~s^{-1}}}\right)^2M_{\odot},
\end{equation}
where we take $\upsilon=\sqrt{3}f\upsilon_{\rm FWHM}/2$ for the corrections of geometry 
and kinetics (Peterson \& Wandel 1999, Fromerth \& Melia 2000, 
McLure \& Dunlop 2001, Krolik 2001).
The factor $f$ remains a matter for debate of BLR geometry,
but $f=1$ is expected. We need to calculate the luminosity at 5100\AA~ from the 
emergent spectrum from the slim disk. 
This has been done in detail by Wang et al. (1999).
We find that $\varepsilon=L_{\rm disk}/L_{5100}=4.0 - 7.5$ for a supermassive black 
hole.  Combining the equations (1) and (2), we have a relation between 
FWHM(H$\beta$) and $M_{\rm BH}$,
\begin{equation}
M_{\rm BH,6}=\left(\frac{{\cal E}_{\rm R}}{{\cal C}}\right)^{\beta/(1-\beta)} 
             \upsilon_{\rm FWHM,3}^{2/(1-\beta)},
\end{equation}
where $\upsilon_{\rm FWHM,3}=\upsilon_{\rm FWHM}/10^3$km~s$^{-1}$ and the constant 
${\cal C}$ is
\begin{equation}
{\cal C}= \frac{\varepsilon}{1.26}\left(\frac{10}{1.46{\cal R}_0f^2}\right)^{1/\beta},
~~~{\rm and}~~~ {\cal E}_{\rm R}=\frac{L_{\rm disk}}{L_{\rm Edd}}.
\end{equation}
This relation clearly relies on
two model-dependent parameters $\varepsilon$ and ${\cal E}_R$, which are determined by the
spectrum and the structure of the super-Eddington accretion disk. Such a dependence can 
thus be used to probe the structures of the super-Eddington accretion disks.

The photon trapping process efficiently lowers the radiated luminosity from the disk,
resulting in a very weak dependence of the radiated luminosity on $\log \dot{M}$.
For a photon trapping accretion flow, ${\cal E}_{\rm R}\approx 1$ (Wang \& Zhou 1999,
Ohsuga et al. 2002), we have the Eddington limit  from equation (3) as
\begin{equation}
M_{\rm BH,6}={\cal C}_1 \upsilon_{\rm FWHM,3}^{2/(1-\beta)},
\end{equation}
where ${\cal C}_1={\cal C}^{\beta/(\beta-1)}$. For the empirical relation of reverberation 
mapping given by Kaspi et al. (2000), $\beta=0.7$ and ${\cal R}_0=32.9$, we have 
a limit relation of $M_{\rm BH,6}={\cal C}_1 \upsilon_{\rm FWHM,3}^{6.67}$, where 
${\cal C}_1=2.9 - 12.62$ for
$\varepsilon =4.0 - 7.5$ and $f=1$.  The Eddington limit lines are shown in Figure 1.
This is a very strong dependence on FWHM(H$\beta$). 
Those objects below this line in the FWHM(H$\beta) - M_{\rm BH}$ plot have
sub-Eddington accretion rates.

On the other hand, the photon bubble instability may operate in a super-Eddington 
accretion disk and the maximum luminosity from the disk reads 
$L_{\rm max}\approx {\cal E}_{\rm R,max}M_{\rm BH,6}^{1/5}L_{\rm Edd}$
where ${\cal E}_{\rm R,max}\approx 300$ for the maximum accretion rate
(Begelman 2002). Such a high luminosity will strongly 
affect the BLR size, equation (3) for the Begelman limit is
changed to 
\begin{equation}
M_{\rm BH,6}={\cal C}_2 \upsilon_{\rm FWHM,3}^{10/(5-6\beta)}
\end{equation}
where ${\cal C}_2=\left({\cal E}_{\rm R, max}/{\cal C}\right)^{5\beta/(5-6\beta)}$. 
The spectrum from a disk with the photon bubble instability is highly unknown, we
take $\varepsilon=4$ to get ${\cal C}_2$ for an illustration, we have
$M_{\rm BH,6}=8.0 \times 10^{12}\upsilon_{\rm FWHM,3}^{12.5}$ if the
Kaspi et al.'s relation is used. This relation is highly different from that 
determined by the photon trapping process.  For a disk 
with the photon bubble instability, the FWHM(H$\beta) - M_{\rm BH}$ relation
is beyond the scope of the frame of Figure 1, we draw ${\cal E}_{\rm R}=2$ line for the 
illustration of the Begelman limit in Figure 1.  A potentially realistic 
situation may be a hybrid structure undergoing the photon trapping and photon
bubble instability in the region between the Eddington and the Begelman lines, where
some objects appear if they have a super-Eddington accretion rate.

The Eddington ratio is usually estimated by $L/L_{\rm Edd}=\xi L_{5100}/L_{\rm Edd}$, 
where $\xi=5 - 9$ (Kaspi et al. 2000, Telfer et al. 2002, Netzer 2003). 
If $L/L_{\rm Edd}>1$, this
object is generally thought to  be a super-Eddington accretor. However, we argue
that $L/L_{\rm Edd}=\xi L_{5100}/L_{\rm Edd}>1$ does not simply imply a 
super-Eddington accretion in practice. If
an object with  $\xi L_{5100}/L_{\rm Edd}>1$ really has a super-Eddington accretion rate,
it should locate on the Eddington limit line (photon trapping), or between the
Begelman and Eddington lines (a hybrid structure of photon trapping
and photon bubble instability). The FWHM(H$\beta) - M_{\rm BH}$ relation can be 
used to reliably test the super-Eddington accretion in AGNs.

It should be noted that the FWHM(H$\beta) - M_{\rm BH}$
relation is very sensitive to $\beta$ due to the indices $2/(1-\beta)$ and 
$10/(5-6\beta)$. 
Any small calibrations on the empirical relation of reverberation mapping
will be important to the limit relation. 
$\beta=0.58$ and ${\cal R}_0=33.1$ were suggested by Netzer (2003)
for the cosmology assumed in the present paper 
$H_0=75~{\rm km~s^{-1}~Mpc^{-1}}$, $\Omega_m=0.3$ and $\Omega_{\Lambda}=0.7$
based on the BCES (bivariate correlated errors and intrinsic scatter) estimator. 
We plot the Eddington limit lines for both of the Kaspi et al's and Netzer's values of
($\beta,{\cal R}_0$), and the Begelman limit line for Kaspi et al's value in Figure 1.

\section{Super-Eddington accretions in AGNs}
It is interesting to search super-Eddington accretions and to test their structures 
in AGNs via the limit relation of FWHM(H$\beta) - M_{\rm BH}$.
We assemble objects from the literatures,

\centerline{\includegraphics[angle=-90,width=8.5cm]{limit_fig1.ps}}
\figcaption{
\footnotesize{
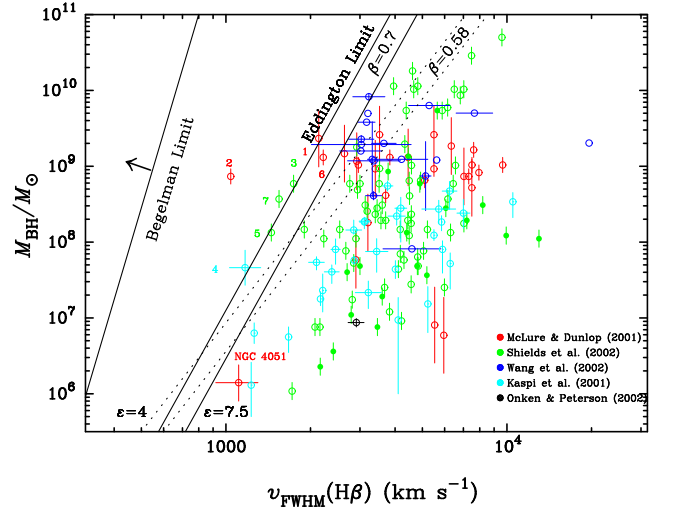
The plot of $\upsilon_{\rm FWHM}({\rm H}\beta) - M_{\rm BH}$.
The dotted and solid lines labelled by the Eddington limit are from equation (5)
for ($\beta=0.58, {\cal R}_0=33.1$) and ($\beta=0.7, {\cal R}_0=32.9$). 
The different color symbols represent the different samples. The open 
circles represent the sources with $L/L_{\rm Edd}<1$ and filled circles 
those with $L/L_{\rm Edd}>1$.  Eddington limit lines (equation 5) are 
labeled by $\varepsilon=4$ and $\varepsilon=7.5$. Begelman limit line 
is drawn for ${\cal E}_R=2$ for illustration.
The objects labelled by numbers are listed in Table 1 for Kaspi et al's 
relation.
\label{fig1}}}
\noindent
in which their black hole masses can
be estimated by 1) the reverberation mapping, 2) the $M_R - M_{\rm BH}$ relation
and 3) the $\sigma - M_{\rm BH}$ relation. The three methods are independent of
the method of the empirical reverberation relation as used by Netzer (2003) in 
a large sample of AGNs. This allows us to test the super-Eddington accretion disks 
in AGNs.  We still estimate the Eddington ratio via
$L/L_{\rm Edd}=\xi L_{5100}/L_{\rm Edd}$, where $\xi=5$ (Telfer et al. 2002, Netzer 
2003) in order to compare with the FWHM(H$\beta) - M_{\rm BH}$ limit relation.

We omitted five objects (IC 4329A, NGC 3227, NGC 7469, PG 1700
and PG 1704) in Kaspi et al.'s sample since their lower uncertainties are too large.
There are 30 objects measured by reverberation mapping, 29 from Kaspi 
et al.'s sample and NGC 3783 from Onken \& Peterson (2002).
McLure \& Dunlop (2001) present 45 objects with host galaxy absolute magnitude
$M_R$, in which 15 Seyfert galaxies of objects 
have been measured by reverberation mapping
(Kaspi et al. 2000). We thus use the results of the reverberation mapping
but we take the absolute magnitudes of IC 4329A, NGC 3227 
and NGC 7469 since their lower limits of the black hole mass are too uncertain
by the reverberation mapping. We take 17 of 37 radio-loud quasars with $M_R$ 
assembled by Wang, Ho \& Staubert (2002)[the rest radio-loud quasars in Wang et al. 
(2002) overlap with that given by McLure \& Dunlop (2001)]. We then 
use the relation
$\log\left(M_{\rm BH}/M_{\odot}\right)=-0.5M_{\rm R}-2.96$ to get the masses
of the black holes (McLure \& Dunlop 2002). Shields et al. (2002) extend the 
relationship $\sigma$$-$FWHM([O {\sc iii}]) in Nelson (2000). Using 
{\rm FWHM}([O {\sc iii}]) as the stellar dispersion velocity $\sigma$, the
relation $M_{\rm BH}=10^{8.13}\left(\sigma/200{\rm km~s^{-1}}\right)^{4.02}M_{\odot}$ 
(Tremaine et al.  2002) enables us to reliably estimate the masses of the black holes  
in 84 quasars (35 RLQs + 49 RQQs) in Shields' sample. 
The error bars in this sample are
given by the largest intrinsic scatter ($<30\%$) (Shields et al. 2002).
Totally there are 164 objects in our present sample. 

Figure 1 shows the $\upsilon_{\rm FWHM}$(H$\beta) - M_{\rm BH}$ plot of the 164 AGNs. 
According to $L/L_{\rm Edd}=\xi L_{5100}/L_{\rm Edd}$ and $\xi=5$,  
19 of 84 quasars in Shields et al.'s sample are super-Eddington accretors. 
The 

\def\mrmbh{$M_R - M_{\rm BH}$}
\def\sigmbh{$\sigma - M_{\rm BH}$}

\begin{center}
\footnotesize
\centerline{\sc  Table 1}
\centerline{\sc List of Candidates of Super-Eddington AGNs}
\scriptsize
\begin{tabular}{llccccl}\hline \hline
    &         &FWHM &\multicolumn{3}{c}{$\log (M_{\rm BH}/M_{\odot})$} & \\ \cline{4-6}
No. &Name     & (H$\beta$)   &\mrmbh & \sigmbh & Rev.& note\\ \hline 
1.  &PG 0157+001 & 2140 & 9.19 & ... & 8.18 & DM01 \\
2.  &Q ~~0204+292 & 1040 & 8.69 & ... & 7.13$^*$ & DM01 \\
3.  &PG 1001+054 & 1740 & ...  &8.77 & 7.65 & S02 \\
4.  &PG 1351+640 & 1170 & ...  & ... & 7.66 & K00 \\
5.  &PG 1440+356 & 1450 & ...  & 8.12& 7.33 & S02 \\
6.  &Q ~~2247+140 & 2220 & 8.94 & ... & 8.07 & DM01\\
7.  &MS 2254$-$37  & 1545 & ...  & 8.17& 7.04 & S02 \\ \hline
\end{tabular}
\vskip 2pt
\parbox{3.2in}
{
\indent
{\sc Notes---}
S02: Shields et al. (2002); MD01: McLure \& Dunlop (2001); K00: Kaspi et al. (2000).\\
$\ast$ there is an error in estimation of 0204+292 in MD01.\\
$\ast\ast$ FWHM(H$\beta$) is in unit of km s$^{-1}$.}
\end{center}
\normalsize

\noindent
values of $\log L/L_{\rm Edd}$ of the 19 quasars
span from zero (VCV 0331-37) to the maximum 0.7 (PG 1216+069),
however these objects are located far below the Eddington limit line as shown by
Figure 1. If we take $\xi=9$, there will be a quite large fraction of the 
objects with $L/L_{\rm Edd}>1$ and most of these objects are still 
located far below the Eddington limit lines. It follows from the Eddington limit
that these objects have sub-Eddington rates 
though they have $L/L_{\rm Edd}>1$. We find that 
PG 1001+054, Mrk 478 and MS 2254-37 with $L/L_{\rm Edd}<1$ are located above the
Eddington limit line. The three may be super-Eddington accretors.
This inconsistency between the Eddington limit line and $\xi L_{5100}/L_{\rm Edd}$
values in some objects reflects: 1) the intrinsic scatter of $\xi$, namely,
the ratio $\xi L_{5100}/L_{\rm Edd}$ is not fully reliable to estimate
$L/L_{\rm Edd}$ as noted by Netzer (2003); 2) the estimation of the black hole mass via 
{\rm FWHM}({\sc [O iii]}) has a larger uncertainty. 
As shown by Boroson (2003) in SDSS EDR sample of 107 radio-quiet quasars and Seyfert 
1 galaxies, the correlation between FWHM([O{\sc iii}]) and $M_{\rm BH}$
is rather scatter, where $M_{\rm BH}$ is determined by Kaspi et al's empirical relation
of reverberation mapping. The FWHM(H$\beta) - M_{\rm BH}$ relation also indicates
the scatters of FWHM([O{\sc iii}])$- M_{\rm BH}$ relation by this inconsistency. 
Therefore,  the FWHM(H$\beta) - M_{\rm BH}$ relation 
may be more reliable to test whether an object has a
super-Eddington accretion rate if the mass of the black hole is well determined.
Mrk 478, PG 1001+054 and MS 2254-37 are narrow line Seyert 1 galaxies 
[FWHM(H$\beta)<2000$~km~s$^{-1}$] (Grupe et al. 1999).
Table 1 lists the estimation of the black hole masses of the three objects.
We find that the differences of $M_{\rm BH}$ given by $\sigma - M_{\rm BH}$ and
the empirical relation of reverberation mapping are large, 1.1, 0.8 and 1.1 dex
for PG 1001+054, Mrk 478 and MS 2254-37, respectively,
showing the mass given by $\sigma - M_{\rm BH}$
is larger than that given by the empirical relation of reverberation mapping.
Even we use the masses by empirical relation of reverberation, the three objects
are still located on or close to the Eddington limit line. We thus suggest that they
are probably super-Eddington accretors.

In the sample of McLure \& Dunlop (2001), we take the data of Mrk 355  
from the reverberation mapping measurements by Kaspi et al.  (2000) and the
data of $M_{\rm BH}=1.4_{-0.6}^{+1.0}\times 10^6M_{\odot}$ and 
FWHM(H$\beta)=1110\pm 190$km s$^{-1}$ in NGC 4051 from Peterson 
et al. (2000).  The values of $L/L_{\rm Edd}$ are given from $L_{5100}$,  
object with $L/L_{\rm Edd}>1$ has not been found in this sample.
But the three radio-quiet quasars PG 0157+001, Q 0204+292 and Q 2247+140 have 
been found above the Eddington limit line. 
Table 1. also lists the three objects. We find that 
the masses of the black holes in PG 0157+001, Q 0204+292 and Q 2247+140
are quite different from $M_R - M_{\rm BH}$ and reverberation relation methods. 
The mass of the black hole in Q 0204+292 is of $10^{8.9}M_{\odot}$ from 
$M_R - M_{\rm BH}$ relation since  the magnitude of its host galaxy 
is quite bright $M_R=-23.3$ (McLure \& Dunlop 2001). On the other hand its 
FWHM(H$\beta)=1040$~km~s$^{-1}$ is very narrow, the black hole mass is 
$10^{7.13}M_{\odot}$. The three objects locate above the Eddington limit lines
even for their masses from empirical reverberation relation.
We suggest that the three objects may be candidates of super-Eddington accretors.

No object with $L/L_{\rm Edd}>1$ has been found in
Kaspi et al.'s sample, but PG 1351+640 locates above Eddington limit line. It may thus
be a super-Eddington accretor. No super-Eddington has been found in the 
radio-loud quasar sample of Wang, Ho \& Staubert (2002), who obtain the Eddington 
ratio from the converting the total emission line luminosity
into ionizing luminosity. There is only one marginal object, PG 2201+315, which is
located close to the Eddington limit whereas the others locate below the Eddington limit.
We would like to stress that the above remarks are based on that the black hole
masses are fully reliable. 

For Kaspi et al.'s relation $\beta=0.7$, we find a handful of objects that locate
above the Eddington limit lines. These objects may be really super-Eddington
accretors. For $\beta=0.58$ empirical relation (Netzer 2003), we find that the 
FWHM(H$\beta) - M_{\rm BH}$ relation becomes flatter 
($M_{\rm BH}\propto \upsilon_{\rm FWHM}^{5.0}$) and there are 23
of 164 ($\approx 14\%$) objects that locate above the Eddington limit lines
(but they are not far away from the Eddington limit lines). If the black hole masses 
of these objects are less uncertain, they are likely super-Eddington accretors. 
We note that these objects locate the region between the Begelman and Eddington 
limit lines. One interesting possibility is that these
objects are undergoing a hybrid process of photon trapping and photon bubble 
instability. It would be necessary to do the observation of
reverberation mapping for these objects or measure the stellar dispersion
velocity of their  host galaxies in order to more reliably
determine the black hole masses and the structure of the super-Eddington
accretion disks. Additionally, the variabilities in the X-ray band will be 
very helpful to understand the physics in these objects (photon trapping 
or photon bubble instability). 

All the objects listed in Table 1 have a narrower width of H$\beta$ 
($<2300$~km~s$^{-1}$). Based on the limit relation, one of prominent features 
of super-Eddington accretion
candidates is the presence of the relatively narrow width of H$\beta$. If
the largest black hole mass is of $M_{\rm BH}^{\rm max}=2\times 10^{10}M_{\odot}$ 
among ANGs (Netzer 2003), the maximum Eddington width is given by
\begin{equation}
V_{\rm FWHM}^{\rm max}({\rm H}\beta)=(3.0 - 3.8)\times 10^3
              \left(\frac{M_{\rm BH}^{\rm max}}{2.0\times 10^{10}M_{\odot}}\right)^{0.15}~~
              {\rm km~s^{-1}}.
\end{equation}
Therefore searching super-Eddington AGNs should be confined 
in those with H$\beta$ width narrower than this Eddington width
$V_{\rm FWHM}^{\rm max}({\rm H}\beta)=3.0 - 3.8 \times 10^3$ km~s$^{-1}$.
An interesting work is to apply the present limits to narrow line 
Seyfert 1 galaxies, which are regarded to be less massive black hole with 
higher accretion rates.  The instability
may make the disk inhomogeneous (Fabian et al. 2002).
However the current data does not allow us to do so.
It is suggested to do measurements of reverberation mapping for a sample of
NLS1 for the determination of the black hole masses 
in order to probe the structures of the super-Eddington accretion disks.

The objects in the present samples used in this paper are low redshift.
The above discussion can be easily extended to the relation between
{\sc C iv}/Mg {\sc ii} width and the mass
of the black hole for high redshift quasars using the empirical
relation of reverberation mapping (Vestergaard 2002, Netzer 2003).

Finally, we would like to point out the response of BLR size to the 
increases of disk's luminosity.
According to the Kaspi et al.'s relation, the virializing timescale of the
clouds in BLR is of 
$\tau_{\rm vir}=\rblr/\upsilon_{\rm FWHM}\approx 7.0M_{\rm BH,8}^{-1/2}
\left(L_{5100}/10^{44}\rm erg~s^{-1}\right)^{1.05}$~(yrs),
where $M_{\rm BH,8}=M_{\rm BH}/10^8M_{\odot}$, which is much shorter than the 
super-Eddington accretion timescale ($\sim 10^6$ yrs). Thus the locations of
H$\beta$-emitting clouds will be
rapidly re-arranged in BLR according to the luminosity. 
If the super-Eddington accretion lasts several million years, we should observe
its effects on the width of broad emission lines. 

\section{Conclusions}

Based on the super-Eddington accretion disk models of the photon trapping 
and photon bubble instability,
we have derived a limit relation between the black hole mass and H$\beta$ width via
the empirical relation of reverberation mapping in AGNs.
We apply the present limits to several samples. We find that most of the objects
locate below the Eddington limit line. In Shields et al.'s sample, there is
a significant fraction of the obejects have $\xi L_{5100}/L_{\rm Edd}>1$, but
they are far below the Eddingtomn limit line. This inconsistency implies that 
the present limit relation is more reliable and  
convenient to test the super-Eddington accretions in AGNs.
We find that there are a handful of objects locate
above the Eddington limit lines, but below the Begelman limit line. We
suggest they may be candidates of super-Eddington accretors. The maximum width 
of H$\beta$ is
$3.0 - 3.8 \times 10^3$~km~s$^{-1}$ for the maximum black holes with
super-Eddington accretion rates among AGNs, which confines the
candidates in future search for super-Eddington accretions among AGNs. 
The FWHM(H$\beta) - M_{\rm BH}$ relation is more reliable and convenient to test 
whether the objects have super-Eddington accretion rates, and can also be used to 
probe the structure of the super-Eddington accretion disks in AGNs.

\acknowledgments{ 
The helpful comments from an anonymous referee are thanked.
I am very grateful to Hagai Netzer for numerous useful suggestions
and comments on the manuscript.
The supports from Alexander von Humboldt Foundation, 
the Special Funds for Major State Basic Research Projects and NSFC are
acknowledged.}

\newpage

\noindent

\end{document}